\documentstyle[12pt,epsfig]{article}
\def\lbldef#1#2{\expandafter\gdef\csname #1\endcsname {#2}}

\def\href#1#2{#2}  

\begin{document}
\baselineskip=15.5pt
\pagestyle{plain}
\setcounter{page}{1}

\begin{titlepage}

\begin{flushright}
CERN-TH/2000-210\\
hep-th/0007114
\end{flushright}
\vspace{10 mm}

\begin{center}
{\Large q-Deformed Conformal Quantum Mechanics}

\vspace{5mm}

\end{center}

\vspace{5 mm}

\begin{center}
{\large Donam Youm\footnote{E-mail: Donam.Youm@cern.ch}}

\vspace{3mm}

Theory Division, CERN, CH-1211, Geneva 23, Switzerland

\end{center}

\vspace{1cm}

\begin{center}
{\large Abstract}
\end{center}

\noindent

We construct a q-deformed version of the conformal quantum mechanics 
model of de Alfaro, Fubini and Furlan for which the deformation parameter 
is complex and the unitary time evolution of the system is preserved.  
We also study differential calculus on the q-deformed quantum phase space 
associated with such system.

\vspace{1cm}
\begin{flushleft}
CERN-TH/2000-210\\
July, 2000
\end{flushleft}
\end{titlepage}
\newpage

\section{Introduction}

It has long been suggested that the description of spacetime based on 
the usual notion of geometry may not be valid at the Planck scale,  
and perhaps the spacetime becomes noncommutative or may show 
non-Archimedean structure at such small length scale.  It has therefore 
been believed that the noncommutative description of spacetime might 
be relevant to quantum theory of gravity.  It is a generic property of 
a noncommutative space that the notion of a point has no meaning but the 
lattice structure of spacetime emerges due to the uncertainty in the 
measurement of the particle  position in space.  So, in some cases  
such lattice structure of spacetime at the Planck scale eliminates 
ultraviolet divergence problem.  Recently, there has been active 
investigation of noncommutative theories after it was found out that 
noncommutative spacetime emerges naturally in M-theory compactified in 
the presence of constant background three-form field \cite{cds} and in 
the worldvolume theory of D-brane with nonzero constant NS $B$-field 
\cite{dh,sew}.  

It is the purpose of this paper to study the noncommutative generalization 
of the conformal quantum mechanics of  de Alfaro, Fubini and Furlan 
\cite{aff}.  In Ref. \cite{cdk}, it is observed that such conformal quantum 
mechanics model can be realized as a special limit of the mechanics of a 
massive charged point particle in the near horizon background of the 
extremal Reissner-Nordstr\" om black hole, indicating the possible relevance 
of the conformal quantum mechanics model to the quantum theory of black 
holes.  Furthermore, the fact that the $SO(1,2)\cong SU(1,1)$ isometry 
symmetry of the AdS$_2\times S^n$ near-horizon geometry of the extremal 
Reissner-Nordstr\" om black hole coincides with the $SL(2,{\bf R})\cong 
SU(1,1)$ symmetry of the conformal quantum mechanics indicates that  
the conformal quantum mechanics may have some relevance to the poorly 
understood AdS$_2/$CFT$_1$ duality.  

In the case of one spatial dimension, it is unclear what is meant by 
noncommutative space (in this paper we assume that the time coordinate 
is a commutative variable), since a coordinate always commutes with itself.  
A possible way of introducing noncommutativity for such case is by following 
the Manin's proposal \cite{man1,man2,man3}, which is based upon a differential 
calculus on quantum plane \cite{wz}, that the Heisenberg algebra can 
be modified through q-deformation of the phase space.   The Manin's proposal 
was first applied to some nonrelativistic dynamical system in Refs. 
\cite{avol,sw}.  Following this line of approach, we shall construct a 
q-deformed version of the conformal quantum mechanics model of Ref. \cite{aff} 
and study its properties.  

The paper is organized as follows.  In section 2, we review the relevant 
aspect of the conformal quantum mechanics.  In section 3, we discuss the 
q-deformed Heisenberg algebra with a complex deformation parameter and 
with unitary time evolution of the system, and apply this to construct a 
q-deformed version of the conformal quantum mechanics.  In section 4, 
we study dynamics of the system and from this we construct differential 
calculus on the q-deformed phase space.

\section{Conformal Quantum Mechanics}

In the following, we briefly summarize the conformal quantum mechanics 
studied in Refs. \cite{aff,ap,fr}.  The Lagrangian density for the system is 
given by
\begin{equation}
{\cal L}={1\over 2}m\dot{x}^2-{g\over{2x^2}}.
\label{harmofcqm}
\end{equation}
The action is invariant under the following $SL(2,{\bf R})$ conformal 
algebra, spanned by the Hamiltonian ${\cal H}$, the dilatation generator 
${\cal D}$ and the special conformal generator ${\cal K}$:
\begin{equation}
[{\cal D},{\cal H}]=2i{\cal H},\ \ \ \  [{\cal D},{\cal K}]=-2i{\cal K},
\ \ \ \   [{\cal H},{\cal K}]=-i{\cal D}.
\label{sl2ralg}
\end{equation}
Here, the $SL(2,{\bf R})$ generators are explicitly given by
\begin{equation}
{\cal H}={{p^2}\over{2m}}+{g\over{2x^2}},\ \ \ \  
{\cal D}={1\over 2}(px+xp),\ \ \ \  
{\cal K}={1\over 2}mx^2.
\label{sl2rgens}
\end{equation}

The problem with the above Hamiltonian ${\cal H}$ is that its eigenspectrum is 
continuous and bounded from below but without an endpoint or ground state, 
and its eigenstates are not normalizable.  Such problem was circumvented 
\cite{aff} by redefining the Hamiltonian as a linear combination of the 
above $SL(2,{\bf R})$ generators with a suitable condition on the 
coefficients.  Particularly, the following choice is found to be 
convenient \cite{aff}:
\begin{equation}
L_0={1\over 2}\left(a{\cal H}+{1\over a}{\cal K}\right),
\label{lzero}
\end{equation}
where the introduction of the constant $a$ leads to breakdown of scale 
invariance.  Then, the potential term in $L_0$ has the minimum and the energy 
eigenstates become discrete and normalizable.  Furthermore, along with the 
following linear combinations:
\begin{equation}
L_{\pm 1}={1\over 2}\left(a{\cal H}-{1\over a}{\cal K}\mp i{\cal D}\right),
\label{lpm}
\end{equation}
$L_0$ satisfies the following $SL(2,{\bf R})$ algebra in the Virasoro form:
\begin{equation}
[L_1,L_{-1}]=2L_0,\ \ \ \ \ \ \ \ 
[L_0,L_{\pm 1}]=\mp L_{\pm 1}.
\label{virasl2}
\end{equation}

\section{q-Deformation of Conformal Quantum Mechanics}

For the ordinary commutative quantum mechanics in two spacetime 
dimensions, the Heisenberg algebra of observables can be defined as the 
following quotient:
\begin{equation}
H(I,x,p)=C[I,x,p]/J(I,x,p),
\label{ordqalg}
\end{equation}
where $C[I,x,p]$ is a unital associative algebra freely generated by 
the identity $I$, the position operator $x$ and the canonical momentum 
operator $p$, and $J(I,x,p)$ is a two-sided ideal in $C[I,x,p]$ generated 
by the following relation corresponding to the Heisenberg rule:
\begin{equation}
xp-px=iI,
\label{heisrul}
\end{equation}
where we are using the unit in which $\hbar=1$.  
The operators $x$ and $p$ are assumed have the following property under 
the antilinear anti-involution operation in $C[I,x,p]$:
\begin{equation}
x^*=x,\ \ \ \ \ p^*=p.
\label{hermcondxp}
\end{equation}  

The formalism of Manin's quantum space \cite{man1,man2,man3} can be applied to 
the above Heisenberg algebra by making use of the q-deformed differential 
calculus developed in Ref. \cite{wz}.  Namely, one can deform the above 
Heisenberg algebra by deforming the usual Heisenberg rule (\ref{heisrul}) 
as follows:
\begin{equation}
xp-qpx=iI,
\label{defheialg}
\end{equation}
where the deformation parameter $q$ can be either complex or real.  
First, if $q$ is a complex number, the consistency of the relation 
(\ref{defheialg}) along with the Hermiticity condition (\ref{hermcondxp}) 
on $x$ and $p$ requires that $|q|=1$.  According to Ref. \cite{avol}, which 
first studied the q-deformed classical and quantum mechanics (with a 
complex deformation parameter $q$) of a particle in one-dimensional space 
and whose work is later generalized to the relativistic case in Ref. 
\cite{rem}, the parameters of the dynamics such as the inertial mass $m$ of 
the particle do not commute with the generators $x$ and $p$ of the algebra 
and there is no unitary time evolution of the system at the quantum level.  
Later, it is found out \cite{brs,rem2,rem3} that to achieve unitary 
noncommutative q-dynamics on the quantum level, i.e. for the Heisenberg 
equation of motion $\dot{\Omega}={i\over\hbar}[{\cal H},\Omega]+\partial_t
\Omega$ to be satisfied after the q-deformation, one has to introduce 
additional generators into the algebra.  Second, if $q$ is a real number, 
$x$ and $p$ cannot both be Hermitian, as can be seen by applying the 
involution operation to Eq. (\ref{defheialg}).  So, one has to assume that 
only one of $p$ and $x$ is Hermitian and the involution of the other is a 
separate operator \cite{sw}.  One can alternatively describe the q-deformed 
Heisenberg algebra with a real $q$ by redefining the generators, say, in 
terms of the above generators $p$, $x$ and $x^*$ so that new momentum and 
position operators can be both hermitian, as was done in Ref. \cite{sw}.  
In such case, an additional generator (expressed in terms of $p$, $x$ and 
$x^*$), which approaches $I$ as $q\to 1$, other than the hermitian position 
and momentum operators, is introduced into the algebra.  Such alternative 
algebra can be obtained \cite{flw} also by making use of the Leibniz rule 
$\partial_{\sf x}{\sf x}=1+q{\sf x}\partial_{\sf x}$ for the differential 
calculus in the one-dimensional q-deformed Euclidean space ${\bf R}^1_q$.
In the present paper, we shall apply the first approach for studying the 
q-deformed generalization of the conformal quantum mechanics of Refs. 
\cite{aff,ap,fr}.  

The q-deformed Heisenberg algebra of observables with a complex deformation 
parameter is given by the following quotient:
\begin{equation}
H=A[I,x,p,K,\Lambda]/J(I,x,p,K,\Lambda).
\label{qdefalg}
\end{equation}
In the case of a particle under the influence of non-trivial potential 
$V$, with the assumption of the proper limit of no q-deformation, the 
two-sided ideal $J$ is defined by the following q-deformed Heisenberg 
relations or the Bethe Ansatz re-ordering rules:
\begin{eqnarray}
xp&=&q^2px+iq\Lambda^2,\ \ \ 
x\Lambda=\xi\Lambda x,\ \ \ \ \ 
p\Lambda=\xi^{-1}\Lambda p,
\cr
xK&=&\xi^{-2}Kx,\ \ \ \ \ 
pK=Kp,\ \ \ \ \ 
\Lambda K=\xi^{-1}K\Lambda.
\label{idecond}
\end{eqnarray}
Here, the generators $K$ and $\Lambda$ are assumed to be invertible 
and time-independent, and one can consistently (with the above q-deformed 
Heisenberg relations) impose the following reality conditions on the 
generators under the involution operation:
\begin{equation}
x^*=x,\ \ \ \ \  p^*=p,\ \ \ \ \  K^*=K,\ \ \ \ \  \Lambda^*=\Lambda,
\label{hermcond}
\end{equation}
along with $|q|=1=|\xi|$.  

In the q-deformed quantum phase space described in the above, the 
Hamiltonian (\ref{sl2rgens}) of the conformal quantum mechanics 
of Ref. \cite{aff} is deformed in the following way:
\begin{equation}
{\cal H}_{\xi}=p^2K^2+{{mg}\over{\xi^4}}x^{-2}K^2\Lambda^4,
\label{qdefharm}
\end{equation}
where we obtained this form of Hamiltonian from the requirement of the 
consistency of the Hamiltonian form of the Heisenberg equations with the 
q-deformed Heisenberg relations (\ref{idecond}), namely the requirement 
of the unitary time evolution of the system.  
To further impose the naturalness condition that the velocity $\dot{x}$ is 
linear in the momentum $p$ in the Heisenberg equation of motion $\dot{x}=
i[{\cal H}_{\xi},x]$, one has to further let $\xi=q$, which we assume from 
now on.  Note, in the limit of no q-deformation, $K$ and $\Lambda$ 
belong to the center of the algebra.  The requirement of irreducibility of 
the representation level implies that they should be proportional to the 
identity $I$ when $\xi=q=1$.  We choose $K={1\over\sqrt{2m}}I$ and 
$\Lambda=I$ when $\xi=q=1$ so that the Hamiltonian (\ref{qdefharm}) reduces 
to the form (\ref{sl2rgens}) in the limit of no q-deformation.  

The dilatation generator ${\cal D}$ and the special conformal generator 
${\cal K}$ in Eq. (\ref{sl2rgens}) of the $SL(2,{\bf R})$ algebra can be 
q-deformed in such a way that the commutation relations (\ref{sl2ralg}) of the 
$SL(2,{\bf R})$ algebra continue to be satisfied after the q-deformation.  
Such q-deformed $SL(2,{\bf R})$ generators are given by
\begin{eqnarray}
{\cal H}_q&=&p^2K^2+{{mg}\over{q^4}}x^{-2}K^2\Lambda^4,
\cr
{\cal D}_q&=&{1\over 2}(qpx+q^{-1}xp)\Lambda^{-2}, 
\cr
{\cal K}_q&=&{1\over{4q^4}}x^2K^{-2}\Lambda^{-4}.
\label{qdefdk}
\end{eqnarray}
By using the q-deformed Heisenberg relations (\ref{idecond}) with $\xi=q$, 
one can show that these q-deformed generators satisfy the following 
commutation relations:
\begin{equation}
[{\cal D}_q,{\cal H}_q]=2i{\cal H}_q,\ \ \ \  
[{\cal D}_q,{\cal K}_q]=-2i{\cal K}_q,\ \ \ \   
[{\cal H}_q,{\cal K}_q]=-i{\cal D}_q.
\label{slq2ralg}
\end{equation}

Just as in the case of undeformed conformal quantum mechanics, one 
can redefine the generators through the linear combinations so that the 
resulting new generators satisfy the $SL(2,{\bf R})$ algebra in the 
Virasoro form.  The q-deformed forms of the $SL(2,{\bf R})$ generators 
(\ref{lzero}) and (\ref{lpm}) are given by
\begin{eqnarray}
L^q_0&=&{1\over 2}\left(a{\cal H}_q+{1\over a}{\cal K}_q\right),
\cr
L^q_{\pm 1}&=&{1\over 2}\left(a{\cal H}_q-{1\over a}{\cal K}_q\mp 
i{\cal D}_q\right),
\label{qdefvirgen}
\end{eqnarray}
where ${\cal H}_q$, ${\cal D}_q$ and ${\cal K}_q$ are defined in Eq. 
(\ref{qdefdk}).  It is straightforward to show that the q-deformed generators 
(\ref{qdefvirgen}) still satisfy Eq. (\ref{virasl2}).  It might be possible 
to construct a q-deformed version of the de Alfaro, Fubini and Furlan 
conformal quantum mechanics in such a way that the generators instead 
satisfy the q-deformed commutation relations and thereby the $SL(2,{\bf R})$ 
algebra (\ref{virasl2}) is deformed to the quantum $SL_q(2,{\bf R})$ algebra. 
The q-deformed version of the symmetry group, the so-called quantum group, 
of a dynamical system was originally studied in Refs. \cite{mac,bie} within 
the context of the quantum noncommutative harmonic oscillator with q-deformed 
creation and annihilation operators.

\section{Differential Calculus on the q-Deformed Phase Space}

Note, the above q-deformed algebra (\ref{qdefalg}) generated by $I$, $x$, 
$p$, $K$ and $\Lambda$, satisfying the commutation relations (\ref{idecond}), 
is the zero-form sector of the q-deformed quantum de Rham complex 
generated by these generators and their differentials.  The quantum 
de Rham complex contains information about not only algebra of 
observables but also dynamics of theory.  Namely, by relating 
the velocity vector $(\dot{x},\dot{p})$ for a particle in the quantum 
phase space to the one forms $dx$ and $dp$ as $dx=\dot{x}dt$ and 
$dp=\dot{p}dt$, one can learn about the dynamics of a particle moving 
on the quantum phase space from the commutation relations among the 
generators of the algebra and their differentials.  Here, the dot denotes 
the derivative with respect to the time coordinate $t$, which we assume 
to be a commuting parameter.  According to Ref. \cite{bdr}, there are 
three families
\footnote{In the case of the relativistic motion of a particle in the 
two-dimensional noncommutative Minkowski spacetime, one of the families is 
excluded and the remaining two coincide \cite{rem} under the condition of 
the reasonable description of the particle dynamics.}
of possible differential calculi associated with the Manin's plane defined 
by the commutation relations among the generators.  In this section, we 
construct the q-deformed quantum de Rham complex directly from the Heisenberg 
equations of motion, instead of applying the result of Refs. \cite{wz,bdr}.  
In the following, we restore the Planck's constant $\hbar$ in the equations 
just for the purpose of making it easy to see various limits.  In particular, 
in the $q\to 1$ and $\hbar\to 0$ limit (i.e., the limit of undeformed 
classical theory), the formulae obtained in the following reduce to those of 
the usual commutative classical geometry.  

The Heisenberg equations associated with the q-deformed Hamiltonian 
(\ref{qdefharm}) with $\xi=q$ have the following form
\footnote{For more general $\xi\neq q$ case, the Heisenberg equations take 
the following form:
\begin{eqnarray}
\dot{x}&=&\left[{i\over\hbar}(\xi^4-q^4)p^2x+q(q^2+\xi^2)p\Lambda^2
\right]K^2, 
\cr
\dot{p}&=&-{{img}\over{\hbar q^4}}\left[\left({q\over\xi}\right)^4-1\right]
px^{-2}K^2\Lambda^4+{{mg}\over{q^4}}\left[\left({q\over\xi}
\right)^2+1\right]x^{-3}K^2\Lambda^6.
\label{genqdefheieqs}
\end{eqnarray}
As mentioned in the previous section, the velocity $\dot{x}$ becomes 
linear in the momentum $p$ when $\xi=q$.}: 
\begin{equation}
\dot{x}={i\over\hbar}[{\cal H}_q,x]=2qpK^2\Lambda^2,\ \ \ \ \ 
\dot{p}={i\over\hbar}[{\cal H}_q,p]={{2mg}\over{q^4}}x^{-3}K^2\Lambda^6.
\label{qdefheieqs}
\end{equation}
One can also show by using Eq. (\ref{idecond}) that $\dot{K}={i\over\hbar}
[{\cal H}_q,K]=0$ and $\dot{\Lambda}={i\over\hbar}[{\cal H}_q,\Lambda]=0$.  
The Aref'eva-Volovich limit \cite{avol} is achieved by further letting 
$\Lambda=I$.   In this case, the system does not evolve unitarily with time, 
as can be seen from the fact that the re-ordering rules (\ref{idecond}), 
which are derived from the condition of unitary time evolution, cannot 
be consistent when $\Lambda=I$.  As expected, in the limit of no 
q-deformation, the above Heisenberg equations of motion reduce to the 
Heisenberg equations associated with the Hamiltonian given by Eq. 
(\ref{sl2rgens}).  

From the above Heisenberg equations, one can express the differentials of 
the generators as follows:
\begin{eqnarray}
dx&=&\dot{x}dt=2qpK^2\Lambda^2dt,\ \ \ \ \ \ 
dp=\dot{p}dt={{2mg}\over{q^4}}x^{-3}K^2\Lambda^6dt,
\cr
dK&=&\dot{K}dt=0,\ \ \ \ \ \ \ \ \ \ \ \ \ \ \ \ \ \,
d\Lambda=\dot{\Lambda}dt=0.
\label{diffgens}
\end{eqnarray}
By using the fact that we defined the time coordinate $t$ to be a commuting 
parameter, one can derive the commutation relations among the generators 
and their differentials.  The following commutation relations can be 
obtained by making use of the relations (\ref{diffgens}) and the q-deformed 
Heisenberg relations (\ref{idecond}) with $\xi=q$:
\begin{eqnarray}
x\,dx&=&dx\,(x+i\hbar q^{-1}p^{-1}\Lambda^2),
\cr
p\,dx&=&q^{-2}dx\,p,
\cr 
K\,dx&=&q^2dx\,K,
\cr 
\Lambda\,dx&=&q^{-1}dx\,\Lambda,
\cr
x\,dp&=&q^2dp\,x,
\cr
p\,dp&=&dp\,(p+3i\hbar qx^{-1}\Lambda^2),
\cr
K\,dp&=&dp\,K,
\cr
\Lambda\,dp&=&qdp\,\Lambda.
\label{derham}
\end{eqnarray}
The first and the sixth relations in Eq. (\ref{derham}) can be 
rewritten in more symmetric forms as follows:
\begin{equation}
px\,dx=q^{-4}dx\,xp,\ \ \ \ 
x^{-3}K^2\Lambda^4p\,dp=q^{-4}dp\,px^{-3}K^2\Lambda^4.
\label{symderham}
\end{equation}
By assuming the usual Leibniz rule and nilpotency condition for the 
external differential operator $d$, one obtains the following product 
rules for the differentials:
\begin{eqnarray}
(dx)^2&=&{{i\hbar}\over 2}q^{-7}p^{-2}\,dxdp,
\cr
(dp)^2&=&{3\over 2}i\hbar qdxdp\,(\partial_xx^{-1})\Lambda^2,
\cr
dxdp&=&-q^2dpdx,
\label{proddiff}
\end{eqnarray}
where $\partial_x$ denotes the generalized q- and $\hbar$-deformed 
partial derivative with respect to $x$, which we define in the following.  
We see that the first commutation relation in Eq. (\ref{derham}) is 
similar to that of the universal calculus over a lattice, except that 
the `lattice spacing' is an operator.  In fact, the q-deformation leads 
to deformation of continuous phase space to a lattice structure, where 
the Hilbert space of representations of the q-deformed system has a 
discrete spectrum, putting physics on a q-lattice.

We further enlarge the algebra by defining the derivatives $\partial_x$ 
and $\partial_p$ on the q-deformed phase space in the following way:
\begin{equation}
d=dx\,\partial_x+dp\,\partial_p,
\label{partialxp}
\end{equation}
along with the assumption of the usual Leibniz rule and the nilpotency 
condition as above.  Note, we have seen in the above that $dK=0=d\Lambda$.  
The following commutation relations between the partial derivatives and the 
generators can be obtained by applying the Leibniz rule:
\begin{eqnarray}
\partial_x\,x&=&1+(x+i\hbar q^{-1}p^{-1}\Lambda^2)\,\partial_x,
\cr
\partial_x\,p&=&q^{-2}p\,\partial_x,
\cr
\partial_x\,K&=&q^2K\,\partial_x,
\cr
\partial_x\,\Lambda&=&q^{-1}\Lambda\,\partial_x,
\cr
\partial_p\,x&=&q^2x\,\partial_p,
\cr
\partial_p\,p&=&1+(p+3i\hbar qx^{-1}\Lambda^2)\,\partial_p,
\cr
\partial_p\,K&=&K\,\partial_p,
\cr
\partial_p\,\Lambda&=&q\Lambda\,\partial_p,
\label{comrelpargen}
\end{eqnarray}
The commutation relations between the partial derivatives and the 
differentials can be obtained by demanding their consistency with the 
q-deformed Heisenberg rules and the product rules obtained in the above.  
We have not yet been successful in obtaining the commutation relations for 
the general $\hbar\neq 0$ case.  In the q-deformed classical phase space 
(the $\hbar=0$ case), the commutation relations are given by
\begin{equation}
\partial_x\,dx=dx\,\partial_x,\ \ \ \ 
\partial_x\,dp=q^{-2}dp\,\partial_x,\ \ \ \ 
\partial_p\,dx=q^2dx\,\partial_p,\ \ \ \ 
\partial_p\,dp=dp\,\partial_p.
\label{comrelpd}
\end{equation}

In the $q\to 1$ and $\hbar\to 0$ limit (i.e., the limit of undeformed 
classical theory), the formulae obtained in the above reduce to those of 
the usual commutative classical geometry.  Particularly interesting limits 
are the $q\to 1$ limit and the $\hbar\to 0$ limit, which respectively 
correspond to the $\hbar$-deformation (or quantization) and the 
q-deformation of the differential calculus on the ``classical'' commutative 
phase space.  Note, the q-deformation and the $\hbar$-deformation generally do 
not commute with one another, i.e., the so-called Faddeev's rectangle is not 
always commutative.  In this paper, we consider the case of the 
q-deformation of the commutative quantized (or $\hbar$-deformed) classical 
conformal mechanics.  Had we first q-deformed the commutative classical 
conformal quantum mechanics and then quantized it, we might have obtained 
different noncommutative theory.

\end{document}